\documentstyle[12pt]{article}

\font\blackboard=msbm10 at 12pt
\font\blackboards=msbm7
\font\blackboardss=msbm5
\newfam\black
\textfont\black=\blackboard
\scriptfont\black=\blackboards
\scriptscriptfont\black=\blackboardss

\def\bfy{{\bf Y}}

\newcommand{\ba}{\begin{array}}
\newcommand{\ea}{\end{array}}
\newcommand{\be}{\begin{equation}}
\newcommand{\ee}{\end{equation}}
\newcommand{\bea}{\begin{eqnarray}}
\newcommand{\eea}{\end{eqnarray}}
\newcommand{\beas}{\begin{eqnarray*}}
\newcommand{\eeas}{\end{eqnarray*}}

\def\half{{1 \over 2}}
\def\identity{{\rlap{1} \hskip 1.6pt \hbox{1}}}

\def\laplace{{\kern1pt\vbox{\hrule height 1.2pt\hbox{\vrule width
1.2pt\hskip
  3pt\vbox{\vskip 6pt}\hskip 3pt\vrule width 0.6pt}\hrule height
  0.6pt}
  \kern1pt}}
\def\scriptlap{{\kern1pt\vbox{\hrule height 0.8pt\hbox{\vrule width
  0.8pt
  \hskip2pt\vbox{\vskip 4pt}\hskip 2pt\vrule width 0.4pt}\hrule height
  0.4pt}
  \kern1pt}}

\def\roughly#1{\raise.3ex\hbox{$#1$\kern-.75em\lower1ex\hbox{$\sim$}}}

\def\tr{{\rm Tr} \,}

\newcommand{\NP}{{\em Nucl.\ Phys.\ }}
\newcommand{\AP}{{\em Ann.\ Phys.\ }}

\newcommand{\PRL}{{\em Phys.\ Rev.\ Lett.\ }}

\newcommand{\gone}[1]{}
\begin{document}
\pagestyle{plain}
\setcounter{page}{1}

\baselineskip16pt

\begin{titlepage}

\begin{flushright}
PUPT-1750\\
hep-th/9712159
\end{flushright}
\vspace{16 mm}

\begin{center}

{\Large \bf Angular momentum\\
and long-range gravitational interactions\\[0.2cm]
in Matrix theory}

\vspace{3mm}

\end{center}

\vspace{8 mm}

\begin{center}

Washington Taylor IV and Mark Van Raamsdonk

\vspace{3mm}
{\small \sl Department of Physics} \\
{\small \sl Joseph Henry Laboratories} \\
{\small \sl Princeton University} \\
{\small \sl Princeton, New Jersey 08544, U.S.A.} \\
{\small \tt wati, mav @princeton.edu}

\end{center}

\vspace{1cm}

\begin{abstract}
We consider subleading terms in the one-loop Matrix theory potential
between a classical membrane state and a supergraviton.  Nontrivial
terms arise at order $v/r^8$ and $v^3/r^8$ which are proportional to
the angular momentum of the membrane state.  The effective potential
for a graviton moving in a boosted Kerr-type metric is computed and
shown to agree precisely with the Matrix theory calculation at leading
order in the long-distance expansion for each power of the graviton
velocity.  This result generalizes to arbitrary order; we show that
terms in the membrane-graviton potential corresponding to
$n$th moments of the membrane stress-energy tensor are reproduced
correctly to all orders in the long-distance expansion by terms of the
form $F^4 X^n$ in the one-loop Matrix theory calculation.
\end{abstract}

\vspace{1cm}
\begin{flushleft}
December 1997
\end{flushleft}
\end{titlepage}
\newpage

\section{Introduction}
\label{Intro}

There is by now abundant evidence that the Matrix theory proposal of
Banks, Fischler, Shenker and Susskind \cite{BFSS} has exposed an
extraordinarily close relationship between supersymmetric matrix
quantum mechanics and 11-dimensional supergravity.  For recent reviews
of the subject, see \cite{banks-review,Susskind-review}.  Numerous
calculations have shown that in various particular cases the leading
term in the one-loop Matrix theory potential reproduces correctly the
leading term in the long-distance supergravity potential (the original
examples of this calculation appeared in \cite{DKPS,BFSS}; a general
proof of this result appears in \cite{Dan-Wati2}).  Subleading terms
in the Matrix theory potential are less well understood, although some
progress has been made towards understanding both subleading terms in
the one-loop potential and terms arising from higher loop effects
\cite{Vijay-Finn2,Becker-Becker,ggr,bbpt,Chepelev-Tseytlin2,Esko-Per2}.
Processes involving longitudinal momentum transfer have been studied
in Matrix theory and found to agree with supergravity
\cite{Polchinski-Pouliot}.  Scattering of more than two gravitons has
also been considered \cite{Dine-Rajaraman}, giving an apparent
discrepancy between Matrix theory and supergravity; a possible
resolution of this discrepancy appears in \cite{Balasubramanian-gl}.
Spin-dependent effects in graviton scattering were considered in
\cite{Harvey-spin,mss,Kraus-spin}.

To date, most studies of subleading effects in Matrix theory have
focused on supergraviton interactions.  The first nonvanishing
subleading term in
the effective potential between a pair of gravitons (without spin) is
of order $v^6/r^{14}$ and arises as a two-loop effect
\cite{bbpt}.  In this paper, we consider
subleading terms in the potential between a classical membrane
configuration of finite size and a single graviton.  The leading term
in the long-distance membrane-graviton potential was calculated for
infinite flat membranes in \cite{Aharony-Berkooz,Lifschytz-Mathur},
and for compact membranes in \cite{Dan-Wati}.  In the case of the
compact membrane it was shown that the potential between the membrane
and a static graviton contains a time-dependent part related to
gravitational radiation from the classically oscillating membrane as
well as a constant part which is proportional to the square of the
energy of the state.  The stationary component of the potential
precisely reproduces the static supergravity potential around a
massive object, so that this result was interpreted as a demonstration
of the equivalence principle in Matrix theory.  Because the membrane
is an extended object, subleading terms arise in this
potential at order $v^k/r^8$ for $k = 0, \ldots, 4$.  In this paper we
calculate these terms, as well as the velocity-dependent terms at
order $1/r^7$, for an arbitrary membrane configuration and analyze
them in detail.  Just as for the static membrane-graviton potential,
we find that each term can be expressed as a sum of a stationary
component and a time-dependent component related to radiation effects.
The stationary components at order $v/r^8$ and $v^3/r^8$ are
proportional to the angular momentum of the membrane state.  We
compare the effective potential calculated in this way to supergravity
using several methods.  We compute the effective potential for a
graviton moving in the  metric produced by a heavily boosted
rotating source in a space which has been compactified in a timelike
direction.  This potential agrees perfectly with the results of our
Matrix theory calculation.  We also compare our results directly with
the potential arising from single-graviton exchange in 11D
supergravity.  We find agreement in this case also.  From the graviton
exchange calculation, it is clear that there are terms of order
$v^k/r^{n+ 7}$ for every $n \geq 0$ in the membrane-graviton potential
which arise from $n$th moments of the stress-energy tensor of the
membrane.  We show that these terms are reproduced precisely to all
orders by certain terms in the one-loop Matrix theory calculation.
 
The paper is organized in the following fashion: In Section 2 we carry
out the one-loop Matrix theory calculation of the membrane-graviton
potential to order $1/r^8$.  We rewrite this potential in the membrane
language of de Wit, Hoppe and Nicolai \cite{dhn} and analyze the
structure of the time-independent components of the potential.  In
Section 3 we calculate the appropriate metric for a boosted rotating
object in lightlike compactified supergravity; we calculate the
effective graviton potential in this background and demonstrate
agreement with large $N$ Matrix theory.  In Section 4 we calculate the
potential from single-graviton exchange processes in supergravity
between a membrane with a given stress-energy tensor and a graviton
with given momentum.  We analyze the effects of higher moments in the
exchange process and show that the resulting terms are reproduced to
all orders in Matrix theory.  Section 5 describes the simple example
of a rotating spherical membrane.  It is found that the usual relation
between angular momentum and the $1/r^8$ term in the potential is
modified by finite $N$ effects.  Section 6 contains concluding
remarks.  Throughout the paper we use the notation and conventions of
\cite{Dan-Wati}.

%
\section{Matrix theory potential at order $1/r^8$}
%

In this section we calculate the one-loop effective potential
governing the long-distance interaction of a graviton with an
arbitrary localized state in Matrix theory.  We arrive at a general
expression for the potential at leading and subleading order in the
inverse separation.  Specializing to states which describe classical
membranes in the large $N$ limit, we simplify the resulting
expressions and show that at each order in $v$ the
potential is described by a time-independent term depending only on
conserved charges of the membrane plus a time-dependent term which
averages to zero. We arrive at a general formula for the time-averaged
potential containing the leading term at each power of the graviton
velocity up to fourth order. For $v$ and $v^3$ the leading
contributions come in at order $1/r^8$ and are proportional to the
transverse angular momentum of the Matrix theory object.

\subsection{One-Loop Matrix theory potential}

The one-loop Matrix theory calculation describing the effective
potential between a pair of separated states was used in \cite{DKPS}
to describe the scattering of a pair of branes in type IIA string
theory.  Since then this calculation has been performed in a wide
variety of contexts, describing interactions between many Matrix
theory objects.  In this subsection we carry out this calculation to
order $1/r^8$ for the potential between a graviton and an arbitrary
Matrix theory object.  We follow the quasi-static approach to this
calculation described in \cite{Dan-Wati}.  It was shown in
\cite{Tafjord-Periwal} that there are discrepancies in some subleading
terms between potentials calculated using this method and those
calculated using the phase shift method of \cite{DKPS}.  However, this
method is known to be valid for the leading order terms, and we find
in this paper that it is also accurate for a particular infinite
series of subleading corrections.

To describe widely separated objects in Matrix theory, one chooses a
background matrix configuration which is block diagonal, with the
trace of each block (divided by the rank) giving the center of mass
coordinates of the subsystem.  We consider a system containing a
compact object with center of mass fixed at the origin and a graviton
having an arbitrary position and velocity. The appropriate background
is given by
\begin{equation}
{\bf X}_i(t) = \left[
\ba{cc}
{\bf Y}_i(t) & 0 \\
0 & r_i(t)
\ea
\right].
\end{equation}
Here, ${\bf Y}_i (t)$ for $i=1,\ldots,9$ are $N \times N$ matrices
solving the classical equations of motion of Matrix theory.  We assume
that $ {\rm Tr}\;({\bf Y}_i (t)) = {\rm Tr}\; (\dot{{\bf Y}}_i (t))=
0$ and that the eigenvalues of ${\bf Y}_i (t)$ have a finite spread so
that we are dealing with a compact object whose center of mass remains
at the origin.  The lower-right entry is a scalar corresponding to a
graviton probe with $p_- = 1/R$.  We assume that the separation
distance $r$ is much greater than the spread of eigenvalues of the
matrices ${\bf Y}_i (t)$, so that our probe is very distant compared
with the extent of our compact object.  We also assume that $v$ is
very small, so that the graviton has moved a short distance compared
to $r$ in the natural time scale associated with the classical
dynamics of the object at the origin.

Using the quasi-static approach of \cite{Dan-Wati}, we expand about
this background in the Matrix theory action and compute the one loop
effective potential.  At quadratic order in the fluctuations of the
off-diagonal degrees of freedom, the action describes a system of
harmonic oscillators with frequencies determined by the background
fields and their time derivatives.  There are 10N complex bosonic
oscillators with $({\rm frequency})^2$ matrix
\beas
\left(\Omega_b\right)^2 &=& M_{0\,b} + M_{1\,b} \\
\noalign{\vskip 0.2 cm}
M_{0\,b} &=& \sum_i K_i^2 \otimes \identity_{10 \times 10} \\
M_{1\,b} &=& \left[ \ba{cc}
0 & - 2  \partial_t K_j \\
2 \partial_t K_i & 2 [K_i,K_j] \ea \right]
\eeas
where
\[
K_i \equiv {\bf Y}_i  -  r_i \;\identity_{N \times N}  \,.
\]
There are also
$16 N\tilde{N}$ complex fermionic oscillators with $({\rm
frequency})^2$ matrix
\beas
\left(\Omega_f\right)^2 &=& M_{0\,f} + M_{1\,f} \\
\noalign{\vskip 0.2 cm}
M_{0\,f} &=& \sum_i K_i^2 \otimes \identity_{16 \times 16} \\
M_{1\,f} &=& i \partial_t K_i \otimes \gamma^i
+ \half [K_i,K_j] \otimes \gamma^{ij}
\eeas
and (two identical sets of) $N\tilde{N}$ complex scalar ghost
oscillators with $({\rm frequency})^2$ matrix
\[
\left(\Omega_g\right)^2 = \sum_i K_i^2\,.
\]
The one
loop Matrix theory potential is  given by
\be
\label{potential}
V_{\rm matrix} = \tr \left(\Omega_b\right) - \half \tr
\left(\Omega_f\right)
- 2 \tr \left(\Omega_g\right)\,.
\ee
For large $r$, $M_0$ and $M_1$ have eigenvalues of order $r^2$ and 1
respectively, so the $M_1$'s can be treated as perturbations when
computing the traces.  For each of the traces, the standard Dyson
perturbation series gives
\bea
\lefteqn{\tr \sqrt{M_0 + M_1} }  \nonumber\\
& = & - {1 \over 2 \sqrt{\pi}} \tr
\int_0^\infty {d \tau \over \tau^{3/2}} \, e^{- \tau (M_0 + M_1)} 
\label{eq:Dyson}\\
& = & - {1 \over 2 \sqrt{\pi}} \tr\sum_n \biggl\lbrace
\int_0^\infty {\prod d \tau_i \over (\sum \tau_i)^{3/2}} \,
e^{-(\tau_1 + \cdots + \tau_{n+1}) r^2} 
e^{-\tau_1 \hat{M}_0} M_1 e^{-\tau_2 \hat{M}_0} \cdots M_1
e^{-\tau_{n+1} \hat{M}_0}\biggr\rbrace \nonumber\\
& = & - {r^{1-2n} \over 2 \sqrt{\pi} } \tr\sum_n \biggl\lbrace
\int_0^\infty {\prod d \sigma_i \over (\sum \sigma_i)^{3/2}} \,
e^{-(\sigma_1 + \cdots + \sigma_{n+1})} 
e^{{-\sigma_1  \over r^2}\hat{M}_0} M_1 e^{{-\sigma_2  \over
r^2}\hat{M}_0} \cdots M_1 e^{{-\sigma_{n+1}  \over 
r^2}\hat{M}_0}\biggr\rbrace\,. \nonumber
\eea
Here, we have taken $\sigma_i = r^2 \tau_i$ and $M_0 = r^2 (\identity
\otimes \identity) + \hat{M}_0 = r^2 (\identity
\otimes \identity) - \tilde{M}_0+ \tilde{\tilde{M}}_0$ where
\[
\tilde{M}_0 =  \sum_i 2r_i {\bf Y}_i \otimes \identity, \;\;\;\;\;
\tilde{\tilde{M}}_0 =  \sum_i {\bf Y}_i^2 \otimes \identity.
\]
Expanding 
\[
e^{-\sigma_i \hat{M}_0/r^2} = 1 - \sigma_i \hat{M}_0/r^2 + \cdots,
\]
we see that a term in the expansion with $n$ powers of $M_1$ and $m$
powers of $\hat{M}_0$ contains terms with powers of $1/r$ ranging from
$1/r^{2n+m-1}$ to $1/r^{2n+2m-1}$.  All terms for $n = 0$, $n = 1$,
$n=2$ and $n=3$ vanish in (\ref{potential}) by Lorentz invariance
\cite{Dan-Wati}, so for a potential calculation to order $1/r^8$, we
need only consider the terms with $n=4,m=0$ and $n=4,m=1$.  Keeping
only these terms we find, to order $1/r^8$
\be
\label{potential2}
V_{\rm matrix} = -\frac{5}{128\;r^7}W_0 - \frac{35}{256\;r^9}W_1
\ee
where
\begin{eqnarray*}
W_0 &=& \tr(M_{1\,b}^4 )- \frac{1}{2} \tr(M_{1\,f}^4)\\
W_1 &=& \tr(\tilde{M}_{0\,b} M_{1\,b}^4) -
\frac{1}{2}\tr(\tilde{M}_{0\,f} M_{1\,f}^4)\,.
\end{eqnarray*}
Computing the traces, we find 
\begin{eqnarray}
W_0 &=& \tr({\cal F})\nonumber\\
W_1 &=& \tr(2r_i {\bf Y}_i \cal F)\
\label{Wterms}
\end{eqnarray}
where
\beas
{\cal F} &=&   
   8 F^\mu{}_\nu F^\nu{}_\lambda F^\lambda{}_\sigma F^\sigma{}_\mu
+ 8( F^\mu{}_\nu F^\lambda{}_\mu F^\nu{}_\sigma F^\sigma{}_\lambda +
   F^\mu{}_\nu F^\lambda{}_\sigma F^\sigma{}_\mu 
F^\nu{}_\lambda)\\
&& \qquad \qquad -  2(F_{\mu\nu} F^{\mu \nu} F_{\lambda \sigma}
   F^{\lambda \sigma} + F_{\mu \nu} F_{\lambda \sigma} F^{\lambda
   \sigma} F^{\mu \nu})
-  2 F_{\mu\nu} F_{\lambda \sigma} F^{\mu\nu} F^{\lambda \sigma} 
 \,.\nonumber
\eeas
The field strength components are given by $F_{0i} = - F_{i0} =
\partial_t {\bf Y}_i - v_i$, $F_{ij} = i [{\bf Y}_i, {\bf Y}_j]$ and
indices are raised and lowered with a Minkowski metric
$\eta_{\mu\nu} = {\rm diag}(- + \cdots +)$.

Note that the only place in the potential that $v_i$ appears is
through $F_{0i}$.  Since the expansion of ${\cal F}$ contains terms
with zero, two and four $F_{0i}$'s, we will get terms with up to four
powers of $v$ for both $1/r^7$ and $1/r^8$.

\subsection{Matrix-membrane correspondance}

We will now restrict attention to systems where the object at the
origin is a membrane.  The correspondence between the Matrix theory
description of a membrane and the world-volume description of a
supergravity membrane in light-front coordinates was developed in
\cite{Goldstone-Hoppe,Hoppe-membrane,bst2,dhn}.  We now review some
details of this correspondence, using the conventions of
\cite{Dan-Wati}.

The essential feature of the matrix-membrane correspondence is that
matrices are taken to correspond to functions on the membrane
world-volume.  The trace becomes an integral, and commutators become
Poisson brackets through
\[
{\rm Tr}\; \rightarrow {N \over 4 \pi} \int d^2 \sigma  \hskip
1.0 cm
 [\cdot,\cdot] \rightarrow
{2 i \over  N}\{\cdot,\cdot\}.
\]
We can translate the field strength components appearing in the Matrix
theory action and one-loop potential into membrane language via the
correspondence
\[
F_{0i} \rightarrow \dot{Y}_i \hskip 1.0 cm F_{ij} \rightarrow - {2 \over N}
\{Y_i,Y_j\}\,.
\]
One notable aspect of this correspondence is that while the algebraic
manipulations available for matrices with commutators are mirrored by
those of functions with Poisson brackets, translating a given matrix
expression into the membrane formalism in this way can drop subleading
terms in the $1/N$ expansion, such as those arising from commutators of
field strength components.  For this reason, we expect results derived
using this formalism to only be valid at leading order in $1/N$.

We now list a number of properties of the membrane variables
which will be used to simplify our expressions.
The transverse coordinates satisfy the equations of motion (equivalent
to
the matrix equations of motion)
\[
\ddot{Y}_i = {4 \over N^2} \partial_a \left(\gamma \gamma^{ab}
\partial_b Y_i\right) \
\]
which follow from the Hamiltonian (equivalent to the matrix
Hamiltonian)
\[
H = {N \over 4 \pi R} \int d^2 \sigma \left( \half \dot{Y}_i \dot{Y}_i
+ {2 \gamma \over N^2} \right). \\
\]
Here, $\gamma_{ab} \equiv \partial_a Y_i \partial_b Y_i$ and $\gamma
\equiv \det \gamma_{ab}$.  This light-front gauge
membrane Hamiltonian is related to the auxiliary $Y^-$ membrane
coordinate by
\[
H = {N \over 4 \pi R} \int d^2 \sigma \left( \dot{Y}^- \right). \\
\]
Since $H$ has a constant value $E$ for  any solution of the equations
of motion, we may write
\[
Y^-(t,\sigma^a) = {R \over N} E t + \xi(t,\sigma^a)
\]
where $E$ is the light front energy of the membrane and $\xi$ is a
fluctuation satisfying $\int d^2\sigma \; \xi = 0$.  Constraints imposed
by the light front gauge choice imply that $\dot{Y}^-$ satisfies
\beas
\label{constraints}
\dot{Y}^- &=& \half \dot{Y}_i \dot{Y}_i + {2 \gamma \over N^2} \\
\partial_a Y^- &=& \dot{Y}_i \partial_a Y_i \\
\ddot{Y}^- &=& {4 \over N^2} \partial_a \left(\gamma \gamma^{ab}
\partial_b Y^-\right) \
\eeas
Finally, we list a few identities used below which can be easily
checked
\begin{eqnarray*}
\{Y_i,Y_j\} \{Y_j,Y_k\} &=& - \gamma \gamma^{ab} \partial_a Y_i
\partial_b Y_k \\
\{Y_i,Y_j\}\{Y_i,Y_j\} &=& 2\gamma\\
\dot{Y}_j\dot{Y}_k - {4 \over N^2} \gamma \gamma^{ab} \partial_a Y_j
\partial_b Y_k &=& {\partial \over \partial t} (Y_j\dot{Y}_k) - {4  
\over N^2} \partial_a(\gamma \gamma^{ab} Y_j \partial_b Y_k)\\
\int d^2 \sigma\left(
Y_i\dot{Y}_j\dot{Y}_k - Y_i{4 \over N^2} \gamma \gamma^{ab} \partial_a
Y_j \partial_b Y_k \right) &=& \int d^2 \sigma\left(\half{\partial
\over \partial t} 
(Y_i\dot{Y}_jY_k+Y_iY_j\dot{Y}_k-\dot{Y}_iY_jY_k) \right)\\
\end{eqnarray*}
The third and fourth relations also hold if we substitute $Y^-$ for any
of the $Y$'s. 

\subsection{Analysis of $1/r^7$ terms}

We begin by analyzing the terms of order $1/r^7$ in the potential
(\ref{Wterms}).  Our goal is to express each term in terms of a
time-independent part plus the world-volume integral of a
time-derivative of a bounded fluctuation.  The term of order $v^0/r^7$
was shown in \cite{Dan-Wati} to be given by
\beas
W_0[v^0] &=&\;  {N \over 4 \pi} \int d^2 \sigma \left[ 24
\; \dot{Y}_i \dot{Y}_i \dot{Y}_j \dot{Y}_j +
{192 \over N^2} \gamma \dot{Y}_i \dot{Y}_i + {384 \over N^4} \gamma^2
- {384 \over N^2} \gamma \gamma^{ab} \dot{Y}_i \bigl(\partial_a
Y_i\bigr)
\dot{Y}_j \bigl(\partial_b Y_j\bigr)  \right]\\
&=& {N \over 4 \pi} \int d^2 \sigma \, \left[
 96 \; \left(\dot{Y}^-\right)^2 - {384 \over
N^2} \gamma \gamma^{ab} \partial_a Y^- \partial_b Y^- \right]\\
&=&  96 \; {R^2 \over N} E^2 + {N \over 4 \pi} \int d^2 \sigma \, \left[
 96 \; {\partial \over \partial t} \left(\xi \dot{\xi}\right) \right]
 \,.
\eeas
This consists of a constant contribution proportional to the square of
the matrix energy, plus a total derivative which vanishes in a time
average, since the fluctuation $\xi$ is a function of bounded
variation. The linear term in $v$ is
\bea
W_0[v^1] &=&\;  {N \over 4 \pi} \int d^2 \sigma \left[ -96
\; v_i \dot{Y}_i \dot{Y}_j \dot{Y}_j -
{384 \over N^2} \gamma v_i \dot{Y}_i + {768 \over N^2} v_i\gamma
\gamma^{ab} \bigl(\partial_a Y_i\bigr)
\dot{Y}_j \bigl(\partial_b Y_j\bigr)  \right] \nonumber\\
&=& {N \over 4 \pi} \int d^2 \sigma \, \left[-192 \; v_i\left(
\dot{Y}_i\dot{Y}^- - {4 \over N^2} \gamma \gamma^{ab} \partial_a Y_i
\partial_b Y^- \right)\right] \label{eq:w01}\\
&=& {N \over 4 \pi} \int d^2 \sigma \, \left[-192 \; v_i {\partial \over
\partial t} \left(Y_i \dot{\xi}\right) \right] . \nonumber
\eea
This term vanishes in a time average, so the stationary
part of the potential has no linear term in the velocity at order
$1/r^7$.  The $v^2$ term at order $r^7$ is
\beas
W_0[v^2] &=&\;  {N \over 4 \pi} \int d^2 \sigma \left[ 48 \;v^2 \dot{Y}_i
\dot{Y}_i +
{192 \over N^2} v^2 \gamma  +  96 \;v_iv_j\dot{Y}_i \dot{Y}_j - {384
\over N^2} v_i v_j \gamma \gamma^{ab}  \partial_a Y_i
\partial_b Y_j  \right]\\
&=& {N \over 4 \pi} \int d^2 \sigma \, \left[96 \;v^2\dot{Y}^- + 96 \;v_i
v_j (\dot{Y}_i\dot{Y}_j - {4 \over N^2} \gamma \gamma^{ab} \partial_a
Y_i \partial_b Y_j )\right]\\
&=&  96 \;v^2 RE + {N \over 4 \pi} \int d^2 \sigma \,
\left[96 \;v_iv_j{\partial \over \partial t}(Y_i \dot{Y}_j) \right] .
\eeas
Here we get a nonvanishing constant term proportional to the matrix
energy of our object plus a term which vanishes in the time-averaged
potential.  

Though we will be mainly interested in the constant piece,
it is interesting to note that the time-varying portion of the
potential for the $v^0$, $v^1$ and $v^2$ terms may be rewritten as
the second time derivative of a quadratic term in the membrane
coordinates ({\it e.g.}, ${\partial^2_t \int Y_iY_j}$ for the $v^2$ term).  
This is exactly the form of the contribution to the
potential that we expect from the graviton's interaction with the
object's quadrupole radiation. We will not discuss this further here.

At third order in $v$, we have
\[
W_0[v^3] = -{N \over 4 \pi} \int d^2 \sigma \, \left[96 \;v^2v_i\dot{Y}_i
\right] = 0\,,
\]
since $\int \dot{Y}_i$ is proportional to the transverse momentum
which we have assumed is zero. Finally, at order $v^4$, we have a 
non-vanishing constant contribution
\[
W_0[v^4] = 24\;v^4N\,.
\]
Note that this term is independent of all properties of the object
except its longitudinal momentum $p_-$.  

Although we have carried out these calculations in the membrane
language, the $v^2$, $v^3$ and $v^4$ terms are simple enough that
analogous relations can be shown without using the membrane
correspondance.  Mimicking the manipulations above with matrices, we
can derive analogous expressions directly in matrix language.
However, the same is not true of the $v^0$ and $v^1$ terms.  As noted
in \cite{Dan-Wati} for the $v^0$ term, the analogous relations in
Matrix theory break down at subleading order in $1/N$, so that the
stationary terms in the potential are not simply related to the
conserved quantities of the Matrix theory Hamiltonian.

To summarize the results of this subsection, we find that at order
$1/r^7$ the stationary part of the potential contains non-vanishing
terms at zeroeth, second and fourth orders in $v$.

\subsection{Analysis of $1/r^8$ terms} 

To find the leading terms linear and cubic in the velocity, we look to
the order $1/r^8$ terms. Starting with the expression for $W_1$ 
(\ref{Wterms}), we have
\bea
W_1[v^1] &=&\;  {N \over 4 \pi} \int d^2 \sigma \left[ -192
\; r_l v_i Y_l\dot{Y}_i \dot{Y}_j \dot{Y}_j -
{768 \over N^2} r_l v_i Y_l \dot{Y}_i \gamma + {1536 \over N^2} r_l
v_i Y_l\gamma \gamma^{ab}  \bigl(\partial_a Y_i\bigr)
\dot{Y}_j \bigl(\partial_b Y_j\bigr)  \right] \nonumber\\
&=& {N \over 4 \pi} \int d^2 \sigma \, \left[-384\; r_l v_i\left(
Y_l\dot{Y}_i\dot{Y}^- - {4 \over N^2} Y_l\gamma \gamma^{ab} \partial_a
Y_i \partial_b Y^- \right)\right] \label{eq:w11}\\
 &=& -192 \;{R^2E \over N} a_{li} r_l
v_i + {N \over 4 \pi} \int d^2
\sigma \, \left[-192 \; r_l v_i {\partial \over \partial t}
\left((Y_l\dot{Y}_i-Y_i\dot{Y}_l) \xi + {RE  
\over N} Y_lY_i 
+Y_lY_i\dot{\xi}\right) \right]. \nonumber
\eea
We find a non-vanishing contribution to the time-averaged potential
proportional to the energy  $E$ and the transverse angular momentum
tensor of the object,
\[
a_{ij} = {N \over 4 \pi R} \int d^2 \sigma \left(Y_i \dot{Y}_j - Y_j
\dot{Y}_i \right) \sim {1 \over R} \tr\left[\bfy_i \dot{\bfy}_j - \bfy_j
\dot{\bfy}_i \right].
\]
Note that this angular momentum is a conserved quantity both in the
membrane theory and in Matrix theory.

The $v^3$ term is simply given by
\beas
W_1[v^3] &=& {N \over 4 \pi} \int d^2 \sigma \,
\left[-192 \;r_lv^2v_iY_l\dot{Y}_i \right] \\
&=& -96 \;Rv^2v_ir_la_{li} + {N \over 4 \pi} \int d^2 \sigma \,
\left[-96 \;v^2v_ir_l{\partial \over \partial 
t}(Y_lY_i)\right]. \\
\eeas
Again, we have a term proportional to the membrane angular momentum
plus a time derivative which vanishes in the average.  Like the terms
of order $v^2/r^7, v^3/r^7$ and $v^4/r^7$, this term is simple enough
to mimic with a matrix calculation so that the analogous result holds
for an arbitrary matrix configuration at any $N$.  The same is not
true of the term (\ref{eq:w11}), which is subject to $1/N$ corrections.

\subsection{Time-averaged potential}

We now consider only the time-independent part of the potential,
performing a time average to eliminate the other terms.  The result of
the preceding calculations is that
to leading order in $1/r$  for each power of velocity  separately, the
membrane-graviton potential is given by
\begin{equation}
\langle V_{\rm matrix} \rangle = -{15R^2E^2 \over 4N}{1 \over r^7} - {15RE
\over 4} {v^2 \over r^7} - {15N\over 16}{v^4\over r^7} \\
+ {105R^2Ea_{li} \over 4N}{r_lv_i \over r^9} + {105Ra_{li} \over 8}
{r_lv_iv^2 \over r^9}
\label{potl}
\end{equation}
Note that the only properties of the object appearing in this
expression are the conserved quantities of the classical membrane
state: the energy, light-front momentum $p_- = N/R$ and the transverse
angular momentum.  As discussed above, the terms of order zero and one
in the velocity are subject to correction at finite $N$; the remaining
terms are correct for any compact object to all orders in $N$.

It is interesting to compare this result to a similar expression which
was derived in \cite{Kraus-spin} for the interaction potential between
gravitons with spin.  In that case, a similar term appeared at order
$v^3/r^8$, but there was no term of order $v/r^8$.  This is
consistent, since the term of this order in (\ref{potl}) is
proportional to the light-front energy $E$, which vanishes for a
graviton with no transverse velocity.

\section{Classical supergravity potential}
\label{sec:metric}

We would like to compare the potential calculation just performed with
the corresponding result from a classical supergravity calculation. If
the Matrix theory conjecture is correct, we would expect to find the
same result, since we are dealing with a process which should be
adequately described by the classical theory.  In this section we
compare the potential computed above to the effective action for a
graviton moving in the appropriate metric on a compactified space.  We
proceed in two steps.  First, we calculate the leading terms in the
stationary long-distance metric of an arbitrary compact object with
fixed momentum in a direction of space-time which has been
compactified on a lightlike circle. This is the appropriate frame for
comparison with Matrix theory.  Then we determine the Lagrangian for a
graviton moving in this background; we read off the potential from
this Lagrangian,  and compare with the matrix expression derived
above.  A similar procedure was used in \cite{Kraus-spin} to study
spin dependence of graviton scattering in Matrix theory.

\subsection{Long-distance metric in lightlike compactified spacetime}

We will now compute the leading terms in the desired stationary
long-distance metric using supergravity.  We assume that the object at
the origin has no net charges other than those associated with
Poincare symmetries, so that we can calculate the metric at large
distances using Einstein's equations in 11 dimensions.

In uncompactified 11 dimensional spacetime in coordinates where the
body's center of mass is fixed at the origin 
and the metric at infinity is Minkowski the long-distance stationary metric
is given by \cite{Landau-Lifshitz-2,Myers-Perry}
\[
G_{\mu\nu} = \eta_{\mu\nu} + h_{\mu\nu}
\]
where
\begin{eqnarray}
h_{00} &=& \frac{CM}{r^8} + O(1/r^{9}) 
\nonumber\\
h_{0I} &=& \frac{9Ca_{IJ}x^J}{2r^{10}} + O(1/r^{10})
\nonumber\\
h_{IJ} &=& \delta_{IJ}\frac{CM}{8r^8} + O(1/r^{9})
\label{flat}
\end{eqnarray}
Here, $M$ is the total energy of the object in its rest frame,
$a_{IJ}$ is the angular momentum tensor, and $C$ is a constant given by  $C = 
64G/ 3\pi^4$.

To make a comparison with Matrix theory, we should consider the metric
of such an object in a lightlike 
compactified theory in a frame where the object has $p_- = N/R$. To
find this, we follow the kinematics of 
Seiberg \cite{Seiberg-DLCQ}.   
Starting with the metric (\ref {flat}), we boost along $x^{10}$
to a frame where $p_{10} = N/R_s$ 
before compactifying $x^{10}$ on a circle of radius $R_s$.  The
appropriate boost parameter satisfies
\[
\gamma M v = \frac{N}{R_s}
\]

To find the metric in the compactified space, we note that all of the
terms in (\ref{flat}) come from the linearized Einstein equations, so
that to this order the compactified metric may be obtained by a
``method of images'', taking (tildes refer to boosted quantities)
\[
h^{comp} _{\mu \nu} = \sum_n \tilde{h}_{\mu \nu} (\tilde{x}^0, \tilde{x}^i,
\tilde{x}^{10} + 2\pi nR_s).
\]
To compute the leading order terms in $1/r$, we need only keep the zeroeth 
Fourier mode of $h$ on the compact circle, so we may average over 
$\tilde{x}_{10}$ to get
\[
h^{comp} _{\mu \nu} = {1 \over 2 \pi R_s} 
\int_{-\infty}^{+\infty}d\tilde{x}_{10}\tilde{h}_{\mu \nu} (\tilde{x}^0, 
\tilde{x}^i,\tilde{x}^{10})\,.
\]
Finally, we perform another boost in the (opposite) $x^{10}$ direction
with boost 
parameter 
\[
\tilde{\gamma} =  \sqrt{\frac{R^2}{2R_s^2} + 1}
\]
which, in the $R_s \to 0$ limit gives the desired form of the metric
for compactification on a lightlike circle of 
radius $R$ with $p_- = N/R$.
The resulting metric (keeping only leading terms) has 
\vspace{-0.25in}
\begin{center}
\begin{minipage}[t]{2in}
\begin{eqnarray}
h_{ij} &=& \frac{5CM^2}{256 \;Nr^7}\delta_{ij} \nonumber\\
h_{+i} &=& \frac{315M^2R\tilde{C}a_{ij}x^j}{1024N^2r^9}\nonumber\\ 
h_{-i} &=& \frac{315\tilde{C}a_{ij}x^j}{512Rr^9}\nonumber
\end{eqnarray}
\end{minipage}
\hspace{0.3in}
\begin{minipage}[t]{2in}
\begin{eqnarray}
h_{++} &=& \frac{45CM^4R^2}{1024 \;N^3r^7}\nonumber\\
\vspace{0.15in}
h_{+-} &=& {35CM^2 \over 512Nr^7} \hspace{1.7in}\label{metric}
\\
\vspace{0.15in}
h_{--} &=& \frac{45CN}{256 \;R^2r^7}\nonumber 
\end{eqnarray}
\end{minipage}
\end{center}
Note that to this order, the components $a_{10 \, i}$ do not appear in
the expressions for $h$.  Both $h_{++}$ and $h_{--}$ have
contributions at order $1/r^8$ proportional to $a_{10 \, i}x^i/r^9$,
but these are at subleading order.


 \subsection{Calculation of graviton potential}
 
We will now compute the effective potential for a graviton moving in a
metric of the type just described.  We follow the approach used by
Becker, Becker, Polchinski and Tseytlin in \cite{bbpt}.  The action
of a scalar particle  in 11-dimensional gravity is
\[
S = - m \int d\tau \,(-G_{\mu\nu}\dot x^\mu \dot x^\nu)^{1/2}
\]
where $G_{\mu\nu}$ is the background.  For the process we will
consider, $p_-$ is to be fixed, so the appropriate Lagrangian in the
transverse coordinates is 
\begin{equation}
{\cal L}'(p_-)\ =\ {\cal L} - p_- \dot x^-(p_-) \ .
\label{routhian}
\end{equation}
Now, from the action, we calculate: 
\[
p_- = m\, \frac{\left(G_{+-} + h_{--} \dot x^- + h_{-i}\dot x^i
\right)}{\left( -G_{\mu\nu}\dot x^\mu \dot x^\nu 
\right)^{1/2}}\ 
\]
where
\[
G_{\mu\nu} = \eta_{\mu\nu} + h_{\mu\nu}\ .
\]
Solving for $\dot x^- (p_-)$ and plugging into (\ref{routhian}) we
find in the $m \to 0$ limit at fixed $p_-$
\begin{equation} 
{\cal L}' = p_- \frac{(1 - h_{+-} - h_{-i}\dot x^i) - \sqrt{(1 -
h_{+-} - h_{-i}\dot x^i)^2 - h_{--} (h_{++} + 
2h_{+i}\dot x^i + v^2 + h_{ij} \dot x^i \dot x^j)} }{h_{--}}\ 
\label{lagrangian}
\end{equation}
In our case, all components of $h$ fall off at least as fast as
$1/r^7$, so up to order $1/r^{13}$ we need only keep 
terms with one power of $h$ in the expansion of
(\ref{lagrangian}). This gives
\begin{equation}
{\cal L}' = p_- \left[\frac{v^2}{2} + \frac{1}{2} h_{++} + h_{+i} v^i
+ \half h_{ij} v^i v^j + \frac{1}{2} h_{+-}v^2 + \frac{1}{2} 
h_{-i} v^2 v^i + \frac{1}{8} h_{--} v^4 + O(1/r^{14})\right].
\label{lagrangian2}
\end{equation}

Each component of the metric appears in this expression coupled to a
different term in the velocity expansion. Thus, the graviton is a
probe of all components of the metric.  We now use the values of $h$
calculated in (\ref{metric}) and read off the effective potential. In
order to compare with the Matrix theory calculation in Section 2, we recall
that the matrix energy of a state is related to its rest frame energy
by
\[
E_{{\rm matrix}} = {R \over 2N} M^2
\]
in the convention with $H_{\rm matrix} = R^{-1} {\rm Tr}\; (\half
\dot{{\bf Y}}^i
\dot{{\bf Y}}^i -  \frac{1}{4}[ {\bf Y}^i, {\bf Y}^j]^2) $.  Also, in
these conventions we have $G = 2 \pi^4 
R^3$, so $C=128R^3/3$. Using these expressions and $p_- = 1/R$, we
find
\[
V_{{\rm gravity}} = -{15R^2E^2 \over 4N}{1 \over r^7} - {15RE \over 4} {v^2
\over r^7} - {15N\over 16}{v^4\over r^7} \\
+ {105R^2Ea_{li} \over 4N}{r_lv_i \over r^9} + {105Ra_{li} \over 8}
{r_lv_iv^2 \over r^9}
\]
This is exactly the stationary part of the potential calculated in
Matrix theory (\ref{potl}). Thus, we have shown that for an arbitrary
compact membrane state in Matrix theory (and in complete generality
for the $v^2$, $v^3$ and $v^4$ terms) the time-averaged one-loop
Matrix theory potential for a distant graviton reproduces the supergravity
result for the leading term in $1/r$ at each power of velocity.

\section{Graviton exchange and higher order terms}


\subsection{Graviton exchange and angular momentum}

Another way to understand the correspondence between the Matrix theory
potential calculation and supergravity is by considering
single-graviton exchange processes in supergravity.  The effective
action from one-graviton exchange between an extended object with
stress-energy tensor $T^{\mu \nu}$ and a
pointlike object with momentum $p^\mu$ located at $y$ in a flat
background metric in light-front coordinates is
\begin{equation}
S_{{\rm eff}} =  - {1 \over 8}\int d^{11} x \, T_{\mu \nu}(x)
D^{\mu\nu\lambda\sigma}(x-y) \frac{p_{\mu} p_{\nu}}{p^+} 
\label{eq:one-graviton}
\end{equation}
where the (harmonic gauge) graviton propagator in $11$ spacetime
dimensions is
\[
D^{\mu\nu\lambda\sigma}(x-y) = 16 \pi G \left(\eta^{\mu\lambda}
\eta^{\nu\sigma} + \eta^{\mu \sigma} \eta^{\nu \lambda} - {2 \over 9}
\eta^{\mu \nu} \eta^{\lambda \sigma} \right) 
\int {d^{11} k \over (2 \pi)^{11}}
{e^{i k \cdot (x-y)} \over - k^2}\,.
\]
The stress tensor for a small object with center of mass at
light-front
coordinates $(z^-,z^i)$ can be expressed in a moment expansion
\[
T^{\mu\nu}(x^+,x^-,x^i) =  T^{\mu \nu} \, \delta(x^--z^-)
\delta(x^i-z^i)\,+
T^{\mu \nu (\lambda)} \partial_{z^\lambda} \, \left( \delta(x^--z^-)
\delta(x^i-z^i)\right)\,+ \cdots
\]
where the zeroeth and first moments of the stress-energy tensor are
given
by 
\begin{eqnarray*}
T^{\mu \nu} & = &  \int d z^- 
d^{9} z^i \; T^{\mu \nu} (z^+, z^-, z^i)\\
T^{\mu \nu (\lambda)} & = &  \int d  z^- 
d^{9} z^i\; [z^\lambda T^{\mu \nu} (z^+, z^-, z^i)].
\end{eqnarray*}
The supergravity interaction potential can be computed exactly to
order $1/r^8$ from these expressions.  For example,
the term linear in the graviton velocity appears in the potential with
terms of
the form
\begin{equation}
V_{{\rm gravity}}[v^1] =
\frac{15 R v^i}{2 \; r^7}  \left[ T^{-i} + \frac{7T^{-i(j)}r_j}{r^2}
+ \cdots\right]. \label{eq:velocity-2}
\end{equation}
To compare this with the Matrix theory calculation in membrane
language we note that for the membrane
\begin{equation}
T_m^{-i} =
\;  {N \over  4 \pi R} \int d^2 \sigma 
\left(
\dot{Y}_i\dot{Y}^- - {4 \over N^2} \gamma \gamma^{ab} \partial_a Y_i
\partial_b Y^- \right).
\label{eq:tm}
\end{equation}
This expression
appeared in the formula (\ref{eq:w01}) for $W_0[v^1]$ and appeared
with an extra factor of $Y_l$ in (\ref{eq:w11}) for $W_1[v^1]$.  Thus,
these terms are precisely proportional to the zeroeth and first
moments of the membrane stress-energy tensor component $T^{-i}_m$.  This
shows that even before time-averaging, Matrix theory correctly
reproduces the expected supergravity potential for these terms.  A
similar argument can be used to show that the terms proportional to other
powers of the velocities are also proportional to moments of the
membrane stress-energy tensor.

The time-dependence of the membrane-graviton effective potential
arises from the fact that components such as $T^{-i}$ of the membrane
stress-energy tensor are not conserved in light-front time.  From the
point of view of supergravity, this time-dependent potential can be
understood in terms of outgoing gravitational radiation which gives
rise to an instantaneous time-dependent potential in light-front
coordinates \cite{Hellerman-Polchinski,Dan-Wati2}.  Thus, in order to
compare to a stationary metric of the type considered in Section
\ref{sec:metric} we must time-average the components of the
stress-energy tensor.  As we saw, this gives a precise agreement
between the Matrix theory calculation and the effective potential in a
static supergravity background metric.  In fact, the structure of the
terms in the potential (\ref{lagrangian2}) arises precisely from the
structure of the one-graviton exchange term (\ref{eq:one-graviton}).
The different components of the metric $h_{\mu \nu}$ are directly
related to the components $T_{\mu \nu}$ of the stress-energy tensor of
the extended object.

\subsection{Higher order terms}

We have seen that the $1/r^8$ terms in the Matrix theory potential
correspond to angular momentum and other first moments of the membrane
stress-energy tensor.  It is natural to ask whether further subleading
terms can be related to higher moments of the stress-energy tensor.
Indeed, this is the case.  We will now prove that all the higher-order
terms proportional to the graviton velocity which arise from single
graviton exchange processes can be reproduced by considering terms in
(\ref{eq:Dyson}) with $n = 4$ and arbitrary $m$.

Generalizing (\ref{eq:velocity-2}) to higher moments, we find that 
\begin{eqnarray}
V_{{\rm gravity}}  &= &
\sum_{p = 0}^{\infty}
\frac{15 R v^i}{2}  \left[  (-1)^p
\frac{1}{p !} 
T^{-i(j_1j_2 \cdots j_p)}\partial_{j_1} \partial_{j_2} \cdots
\partial_{j_p} (\frac{1}{r^7})
\right]\nonumber\\
& = &
\sum_{p = 0}^{\infty}
\frac{R v^i}{2 \; r^{7 + p}}  \left[ 
\sum_{k  \leq p/2}  (-1)^{k}
T^{-i(j_1j_2 \cdots j_p)} (\eta_{j_p j_{p -1}} \eta_{j_{p-2}j_{p-3}}
\cdots \eta_{j_{p-2k + 2}j_{p-2k +1}}) \right. \label{eq:velocity-all}\\
& &\hspace{1.7in} \left.\times \frac{(5 + 2p-2k)!! \; r_{j_1}r_{j_2} \cdots
r_{j_{p-2k}}}{2^k\;k!  \;(p-2k)!  \;r^{p-2k}} 
\right] \nonumber
\end{eqnarray}
where $n!!= n (n-2) (n-4) \cdots 1$.  

We can compare this with the higher order terms in the Matrix theory
potential arising from $n = 4$ contributions to (\ref{eq:Dyson}).  In
the matrix membrane language, each power of $\tilde{M}_0/r^2$ which enters
contributes a factor of $-r\cdot Y/r^2$ while each factor
of $\tilde{\tilde{M}}_0/r^2$ contributes $Y^2/r^2$.  Because on the
membrane world-volume the functions $Y$ are commuting, it does not
matter at which position in the trace these terms contribute.  It is
relatively straightforward to analyze the combinatorial structure of
the various possible terms.  We can compute
\begin{eqnarray*}
\lefteqn{ - {1 \over 2 \sqrt{\pi}  r^7}   \biggl\lbrace
\int_0^\infty {d^5 \sigma_i \over \sigma^{3/2}} \,
e^{-\sigma} 
e^{{-\sigma  \over r^2} (\tilde{M}_0 + \tilde{\tilde{M}}_0)}
\biggr\rbrace}\nonumber \\
 & = &-  \sum_{p = 0}^{\infty} \sum_{k \leq p/2}
 {(-1)^{p+k} \over 2 \sqrt{\pi}   r^{7 + 2p-2k}}  \biggl\lbrace
\int_0^\infty {d^5 \sigma_i \over \sigma^{3/2}} \,
e^{-\sigma} \sigma^{p-k}\frac{1}{(p-k)!} \left(\begin{array}{c}
p-k \\k
\end{array} \right)
\tilde{M}_0^{p-2k} \tilde{ \tilde{M}}_0^{k}
\biggr\rbrace
\nonumber \\
 & = & \sum_{p = 0}^{\infty} \sum_{k \leq p/2}
\frac{ (-1)^{k + 1}\; (5 + 2 p-2k)!!}{3 \cdot 2^{7 + k} \; k! \; (p-2k)!
\; \; r^{7 +2p-2k}} \; (r \cdot {\bf Y})^{p-2k} {\bf Y}^{2k} 
\end{eqnarray*}
where we have abbreviated $\sigma = \sigma_1 + \cdots + \sigma_5$.
{}From this and the fact that
$W_0[v^1] = 192 \; R T^{-i}_m v^i$ it follows that
\begin{eqnarray*}
V_{{\rm matrix}}[v^1] &= &
  \sum_{p = 0}^{\infty} \frac{R v^i}{2 \; r^{7 + p}}
\sum_{k \leq p/2}
\left[(-1)^{k}
T_m^{-i(j_1j_2 \cdots j_p)} (\eta_{j_p j_{p -1}} \eta_{j_{p-2}j_{p-3}}
\cdots \eta_{j_{p-2k + 2}j_{p-2k +1}}) \right. \\
& &\hspace{1.7in} \left.\times \frac{(5 + 2p-2k)!! \; r_{j_1}r_{j_2} \cdots
r_{j_{p-2k}}}{2^k\;k!  \;(p-2k)!  \;r^{p-2k}} 
\right]
\end{eqnarray*}
where we define the higher moments $T_m^{-i(j_1 \cdots j_p)}$
of the membrane stress tensor through (\ref{eq:tm}) with  the product
$Y^{j_1} \cdots Y^{j_p}$ inserted into the integral.  This expression
agrees precisely with (\ref{eq:velocity-all}).  Thus, we see that
there is an exact agreement at all orders between the set of terms in
the one-loop Matrix theory potential which are linear in the velocity
and cubic in the membrane field strength and the terms in the
supergravity potential which are linear in the velocity and arise from
single-graviton exchange processes.  It is straightforward to
generalize this argument to the terms proportional to all powers in
the velocity from zero through four.

\section{Example: Rotating spherical membrane}

As an explicit example of how transverse angular momentum appears in
the Matrix theory potential at order $v/r^8$, we consider a symmetric
spherical membrane of radius $\tilde{R}$ initially at
$x_1^2+x_2^2+x_3^2 = \tilde{R}^2$ and rotating uniformly in the $1-4$
plane, the $2-5$ plane and the $3-6$ plane.  Such a spherical membrane
can be described by matrices which are linear in the $N \times N$
$SU(2)$ generators in $U(N)$ through
\beas
Y_i &=& {2\tilde{R} \over N} J_i \cos(\omega t) \\
Y_{i+3} &=& -{2\tilde{R} \over N} J_i \sin(\omega t)\\ 
Y_{7,8,9} &=&  0
\eeas
where the $N \times N$ matrices $J_i$ satisfy
\[
[J_i,J_j] = i\epsilon_{ijk}J_k\,.
\]
It is easily verified that this configuration solves the equations of motion 
\[
\ddot{Y}_i + [[Y_i,Y_j],Y_j]=0
\]
for $\omega = 2\sqrt{2}\tilde{R}/N$. At this frequency, the
centrifugal forces are sufficient to keep the membrane from
collapsing, and the rotating sphere has a constant radius. The angular
momentum of this state is
\beas
a_{ij} &=& {1 \over R} \tr [Y_i\dot{Y}_j-Y_j\dot{Y}_i]\\
&=&{2\sqrt{2}\tilde{R}^3 \over 3R} (1-{1\over N^2})c_{ij}
\eeas
where $c_{41}=-c_{14}=c_{52}=-c_{25}=c_{63}=-c_{36}=1$ and the matrix energy 
is
\beas
E &=& {1 \over R} \tr\left(\half\dot{Y_i}\dot{Y_i}- {1 \over 4} 
[Y_i,Y_j][Y_i,Y_j]\right)\\
&=& {6\tilde{R}^4 \over  RN}(1-{1 \over N^2})\ .
\eeas
Let us now consider the interaction between this rotating sphere and a
graviton with position $r_i$ and velocity $v_i = \dot{r}_i$.
The term in the Matrix theory potential proportional to $v/r^8$ is given by 
equation (\ref{Wterms}) as
\beas
V_{v/r^8} &=& {35 \over 256 r^9} \tr[2r_iY_i{\cal F}_{v^1}]\\
 &=&{3360\sqrt{2} \tilde{R}^7 \over r^9N^7}
\left(v_i \sin(\omega t) + v_{i+3}
\cos(\omega t)\right) \left(r_l \cos(\omega t) -r_{l+3}\sin(\omega
t)\right)\\
& &\hspace{1.2in} \times\tr[J_l(J_iJ_jJ_j+J_jJ_jJ_i+J_j J_i J_j)]\ . \\
\eeas
The $\cos(\omega t)\sin(\omega t)$ terms time-average to zero while the 
$\cos^2$ and $\sin^2$ terms average to $1/2$, so, computing the trace, the time 
averaged potential becomes
\[
\langle V_{v/r^8} \rangle = {105 \sqrt{2} \tilde{R}^7 \over r^9 N^2}
(r_i v_{i+3} -  
r_{i+3}v_i)\left(1-{10 \over 3N^2} +{7 \over 3N^4}\right)\ .
\]
Using the explicit expressions for $E$ and $a_{ij}$ for this state, we have 
\[
- {105R^2Ea_{li} \over 4N}{r_lv_i \over r^9} = {105 \sqrt{2}
\tilde{R}^7 \over r^9  
N^2}(r_i v_{i+3} - r_{i+3}v_i)\left(1-{2 \over N^2} +{1 \over
N^4}\right)\ .
\]
Thus, we see explicitly that in the large $N$ limit, 
\[
\langle V_{v/r^8} \rangle = - {105R^2Ea_{li} \over 4N}{r_lv_i \over r^9}
\]
for this state, and so the matrix energy and angular momentum appear
as expected. However, this formula does not hold at finite $N$, since
there is agreement only at leading order in $N$.

\section{Conclusions}

We have analyzed subleading terms in the one-loop Matrix theory
potential between a classical membrane configuration and a moving
graviton.  We found that the terms of order $v/r^8$
and $v^3/r^8$ contained stationary components proportional to the
angular momentum of the membrane state.  These terms were shown to
correspond precisely to the predictions of lightlike compactified
supergravity.  We also showed that the one-loop Matrix theory
potential contains an infinite series of terms of order $1/r^{7+n}$
related to higher moments of the membrane stress-energy tensor, and
that these terms agree precisely with terms in the supergravity
effective potential arising from single graviton exchange.

These results provide further evidence that there is a remarkably deep
structure hidden in Matrix theory which is capable of reproducing
extremely nontrivial properties of 11-dimensional supergravity.  The
methods used here provide a systematic framework with which to further
explore this structure.

We have not discussed finite $N$ effects much in this paper; indeed,
most of our Matrix theory calculations were performed in the membrane
language where $1/N$ corrections are dropped.  There are currently
some apparent contradictions between Matrix theory and supergravity
\cite{dos,ggr,Dine-Rajaraman,Douglas-Ooguri,Dan-Wati,Esko-Per-short},
most of which seem to relate to the validity of the finite $N$ DLCQ
conjecture \cite{Susskind-DLCQ}.  Although there are some fairly
convincing arguments that finite $N$ Matrix theory reproduces DLCQ
M-theory \cite{Sen,Seiberg-DLCQ,dealwis-DLCQ,Balasubramanian-gl},
there are subtleties in these arguments indicating that DLCQ M-theory
may not be equivalent to DLCQ supergravity
\cite{banks-review,Hellerman-Polchinski}.  It was shown in
\cite{Dan-Wati} that the stationary part of the leading term in the
potential between a membrane and a graviton with no transverse
velocity is proportional the square of the Matrix theory energy of the
membrane.  However, it was pointed out that this relationship only
holds at leading order in $1/N$.  There are subleading corrections
which seem to indicate a breakdown of the equivalence principle at
finite $N$.  The manipulations we used to show that the $v/r^8$ term
is proportional to the matrix angular momentum used the membrane
language in which subleading terms in $1/N$ are dropped.  Analogous to
the subleading corrections in the $v^0/r^7$ term, we found subleading
corrections in the $v/r^8$ term which break the relationship between
the coefficient of this term and the membrane angular momentum.  This
gives further evidence that at finite $N$ the usual relationships
between the long-distance gravitational field around a finite-size
object and the conserved quantities of the object break down.  It is
difficult to see how to reconcile this observation with the finite $N$
DLCQ conjecture.

\section*{Acknowledgements}

We are particularly grateful to to Dan Kabat, who helped develop some
of the methods used in this paper, for helpful conversations.
The work of MVR is supported in part by the Natural Sciences and Engineering 
Research Council of Canada (NSERC). The work of WT is supported in part by the 
National Science Foundation (NSF) under contract PHY96-00258.

\bibliographystyle{plain}

\end{document}